\documentclass[aps,prl,superscriptaddress,reprint]{revtex4-1}
\usepackage{graphicx}
\usepackage{amssymb,amsmath}
\usepackage[colorlinks]{hyperref}
\usepackage[small,bf]{caption}
\usepackage{subfig}

\usepackage{float}
\usepackage{color}

\begin{document}

\title{\textbf{Terahertz Harmonic Generation in Graphene.}}

\author{Samwel Sekwao}
%\email[]{ssekwao2@illinois.edu}
\affiliation{Department of Physics and Beckman Institute, University of Illinois, Urbana-Champaign, Urbana, Illinois, 61801}

\author{Jean-Pierre Leburton}
\email[]{jleburto@illinois.edu}
\affiliation{Department of Physics and Beckman Institute, University of Illinois, Urbana-Champaign, Urbana, Illinois, 61801}
\affiliation{Department of Electrical and Computer Engineering, University of Illinois, Urbana-Champaign, Urbana, Illinois, 61801}

\begin{abstract}
\textbf{We show that charge carrier transport in graphene exhibit sharp resonances in the presence of spatially and temporarily modulated scattering. Resonances occur when the period of an applied a-c field corresponds to the time taken by quasi-ballistic carriers to drift over a spatial scattering period, provided the latter is shorter than the distance taken by carriers to emit an optic phonon. We show that such system can be achieved with interdigitated gates energized with an a-c bias on graphene layers. Gate separation and fields to achieve ballistic transport would result in resonances in the terahertz range, with the generation of higher harmonics characterized by large $Q$-factors, which are tunable with gate spacing.}
\end{abstract}
\maketitle %\maketitle must follow title, authors, abstract and \pacs
%\section{Introduction}
As a one atom thick, two-dimensional (2D) material made up of carbon atoms arranged in a honeycomb structure,\cite{Novoselov,Novoselov2} graphene is characterized by linear dispersion relation between carrier energy and momentum $E = \hbar v_F|k|$  at $K$ and $K’$ symmetry points of the 2D Brillouin zone. As a consequence, all charge carriers in graphene move with the constant Fermi velocity $v_F \sim 10^8$ cm/s.\cite{Katsnelson,Saito} The interaction of these carriers with the 2D environment such as lattice vibrations or oscillatory electro-magnetic fields, is of fundamental interest since it opens a new way to study quantum electrodynamics of charged particles in solid state physics.\cite{Mikhailov} It may also have important implications in semiconductor technology.\cite{Vafek,Ryzhii,Vasko} Owing to their large velocity, the weak interaction between charge carriers and acoustic phonons (APs) results in long mean free paths for the former that experience quasi-ballistic motion over distance of the order of several micrometers.\cite{Piscanec,Ferrari} At high energy however, carriers lose their momentum (and energy) to efficient optic phonons (OPs) that have much higher vibration frequency ($\hbar \omega_{OP} \sim 0.2$eV) than in conventional semiconductors.\cite{Tiwari} Therefore, in the presence of a constant electric field $F_o$, the carrier motion results in a succession of quasi-ballistic acceleration (when their energy $E < \hbar \omega_{OP}$ ) and scattering  (when their energy is $E \geq \hbar \omega_{OP}$).\cite{Shockley} This stop-and-go motion of carriers can occur over long distances during picosecond flight times.\cite{Sekwao} An interesting effect could arise if the carriers are placed in periodic long range and time varying scattering to achieve transport resonance as well as possible frequency mixing. This kind of situation could be realized in free standing graphene sheets lying over periodically spaced narrow electric gates that would be regulated by an a-c field of appropriate frequency to modulate coulomb scattering of remote oxide impurities when carriers pass in front of the gates (Fig. 1). In the device, a DC field $F_o$ is set up between the source $S$ and the drain $D$, and the a-c field $F_1$ is applied between successive gates, so that charge carriers experience periodic electric fields and scattering times varying in time and distance along the channel.
\par
In this letter, we show that electronic current in graphene under the influence of time and space dependent periodic scattering and electric fields exhibits sharp resonances in the terahertz range, associated with the generation of higher harmonics characterized by large Q-factors tunable with scattering periodicity. The electron dynamics is investigated with a semi-classical Boltzmann formalism, where the effects of electron-electron interactions are ignored for low carrier concentrations ($n \ll 10^{12}$/cm$^2$).\cite{Castro}
%\setcounter{equation}{1}
%%\section{Method}
\par
We consider electrons in the conduction band of graphene in a field of the form $F(x,t) = F_o + F_1e^{i(qx - \omega t)}$ with $0 < F_1/F_o < 1$, where $F_o$, and $F_1$ are the DC and a-c components respectively. The parameter $q$ is the electric field wave number, and $\omega$ is the a-c frequency. In such fields, electrons accelerate until they gain enough energy to interact with OPs once $E \geq \hbar \omega_{OP}$, and suddenly lose their energy and momentum to re-accelerate quasi-ballistically in the fields, as this process repeats itself. In this dynamical picture, the momentum space can be divided into two regions $I$, and $II$ that corresponds to electron energies $E < \hbar \omega_{OP}$, and $E \geq \hbar \omega_{OP}$ respectively. In region $I$, electrons accelerate while interacting with low energy scattering agents (APs and impurities).
\begin{figure}[htp]
\centering
\includegraphics[width=0.95\linewidth]{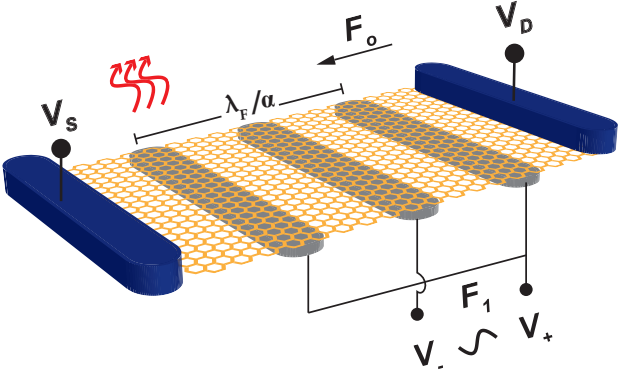}
\caption{Color online) Schematics of a graphene based FET with periodic gating. The DC field $F_o$ is applied between the source and the drain, and the AC field $F_1$ is applied between successive gates. The wavelength of the AC field is twice the distance between successive gates.}
\vspace{-1.0em}
\label{fig1b}
\end{figure}
In region $II$, the electrons interact with OPs and scatter back to region $I$. In low electron concentration ($n < 10^{12}$/cm$^2$), most of the electron population is stretched along the fields (streaming case), so that the majority of carriers have their velocity pointing in that direction.\cite{Sekwao} If the fields are applied along the $x$- direction, the time and space dependent BTE in both regions reads
\begin{align*}\frac{\partial f_I}{\partial t} + v_F\frac{\partial f_I}{\partial x} + \frac{eF(x,t)}{\hbar}\frac{\partial f_I}{\partial k_x} & = -\frac{f_I - f_{FD}}{\tau (x,t)} \\ & + \sum_{\vec k'}S_{OP}(\vec k',\vec k)f_{II}(\vec k',x,t)
\tag{1a}\end{align*}
\begin{align*}
\frac{\partial f_{II}}{\partial t} + v_F\frac{\partial f_{II}}{\partial x} + \frac{eF(x,t)}{\hbar}\frac{\partial f_{II}}{\partial k_x} = -f_{II}\sum_{\vec k'}S_{OP}(\vec k,\vec k')
\tag{1b}\end{align*}
\setcounter{equation}{1}
where $f_I$, and $f_{II}$ are the distribution functions in regions $I$ and $II$ respectively, and $f_{FD}$ is the Fermi-Dirac equilibrium distribution function ($k_F > 0$). Also, in the streaming case, we assume $v_x \sim v_F$. Low energy elastic collisions in region $I$ with impurities and APs are accounted for within a local and time dependent relaxation time approximation\cite{Sekwao}$ \tau(x,t)$ with the same periodicity as the gate spacing and applied a-c field i.e.
\begin{equation}
\frac{1}{\tau(x,t)} = \frac{1}{\tau_o} + \frac{1}{\tau_1}e^{i(qx - \omega t)}
\end{equation}
where $\tau_o$ is a constant relaxation time used to modulate the strength of low energy scattering (with impurities and APs), and $\tau_1$ modulates scattering due to the periodic gates (see Fig 1). The expression $S_{OP}(\vec k, \vec k')$ is the OP transition rate between the states $\vec k$ and $\vec k'$ given by \cite{Lazzeri}
\begin{equation}
S_{OP}(\vec k, \vec k') = \frac{\pi {D_o}^2(N_q + 1)(1 + \cos(\theta'))}{\sigma A \omega_{OP}}\delta(E' - E + \hbar \omega_{OP})
\end{equation}
where $D_o$ is the optical deformation potential, $N_q$ is the phonon occupation number, $\theta'$ is the angle between $\vec k$ and $\vec k'$, $\sigma$ is the area density of the graphene sheet, and $A$ is the area of the sheet. We neglect the effects of optic phonon absorption since their population is negligible, with $\hbar \omega_{OP} \gg k_B T$ even at room temperature. The second term on the Right Hand Side (RHS) of Eq. 1a is due to carriers repopulating region $I$ after scattering with OPs, and the RHS of Eq. 1b is the corresponding depopulation term.
\par
Due to the nature of the problem, we propose a solution of the form
\begin{equation}
f(\vec k,x,t) = f^h(\vec k) + \sum_{m = -\infty}^{\infty}\sum_{n = -\infty}^{\infty}f^{mn}(\vec k) e^{i(mqx - n\omega t)}
\end{equation}
where $ f^h(\vec k) = f^{00}$ is the solution to the homogeneous, steady state problem,\cite{Sekwao} $f^{mn}(\vec k)$ are the harmonic amplitudes, and the summation excludes all terms with $m = 0$ and $n = 0$, since they would result either in homogenous ($m = 0$) or state-state ($n = 0$) terms that are already taken into account in $f^h(\vec k) = f^{00}(\vec k)$. On substituting Eq. 4 into Eq. 1, we get the following recurrent equations for the individual harmonics;
\begin{align*}
\frac{eF_o}{\hbar}\frac{\partial f_I^{mn}}{\partial k_x} & = -\Big[\frac{1}{\tau_o} + i(mqv_F - n\omega) \Big]f_I^{mn}\\& -\frac{eF_1}{\hbar}\frac{\partial f_I^{(m-1)(n-1)}}{\partial k_x} - \frac{f_I^{(m-1)(n-1)}}{\tau_1} \\ & + \sum_{\vec k'}S_{OP}(\vec k',\vec k)f_{II}^{mn}(\vec k')
\tag{5a}\end{align*}
\begin{align*}
\frac{eF_o}{\hbar}\frac{\partial f_{II}^{mn}}{\partial k_x} & = -\Big[\sum_{\vec k'}S_{OP}(\vec k,\vec k') + i(mqv_F - n\omega) \Big]f_{II}^{mn}\\& -\frac{eF_1}{\hbar}\frac{\partial f_{II}^{(m-1)(n-1)}}{\partial k_x}
\tag{5b}\end{align*}
where $f_I^{mn}$, and $f_{II}^{mn}$ are the harmonic amplitudes in regions $I$ and $II$ respectively. On setting $m = n = 0$ into Eq. 5, the terms in $f^{00}$ satisfy the steady state, homogenous Boltzmann equation in the two regions, so that we get the following equations for $f^{-1,-1}$:
\begin{align*}
\frac{eF_1}{\hbar}\frac{\partial f_I^{-1,-1}}{\partial k_x} = - \frac{f_I^{-1,-1}}{\tau_1}\tag{6a}
\end{align*}
\begin{align*}
\frac{eF_1}{\hbar}\frac{\partial f_{II}^{-1,-1}}{\partial k_x} = 0.\tag{6b}
\end{align*}
\par
Eq. 6b has the general solution $f_{II}^{-1,-1}(k_x,k_y) = g(k_y)$ which is unphysical, and should be set to zero, which by continuity in region $I$ also yields $f_I^{-1,-1} = 0$. Because of the recurrence between $f^{mn}$ and $f^{(m-1)(n-1)}$ in Eq. (5), we also have $f^{mm} = 0$ for all $m < -1$.
\par
Let us now consider the harmonics $f^{1n}$ with $n > 1$. Since the R.H.S of Eqn 5b is zero, the solution for $f_{II}^{1n}$ is readily obtained and reads
\setcounter{equation}{6}
\begin{equation}
f_{II}^{1n} = f_{b}^{1n}\exp\Bigg\{-\frac{\hbar}{eF_o}\int_{k_x^o}^{k_x}\Big[\sum_{\vec k'}S_{OP}(\vec k,\vec k')|_{k_x = z} + i(qv_F - n\omega) \Big]dz\Bigg\}
\end{equation}
where the function $f_{b}^{1n}$ is the solution $f_{II}^{1n}$ evaluated at the boundary $k_x = k_x^o = \sqrt{(k_c)^2 - (k_y)^2}$. On substituting Eq. 7 into Eq. 5a, and evaluating the resulting equation at the boundary leads to a homogeneous integral equation of the form,
\begin{equation}
f_{b}^{1n}(k_y) = \int_{-k_x^o}^{k_x^o}\,dyf_{b}^{1n}(y)K(k_y,y,n),
\end{equation}
where $K$ is a complex kernel.
\begin{figure}[h!]
\centering
\includegraphics[width=0.9\linewidth]{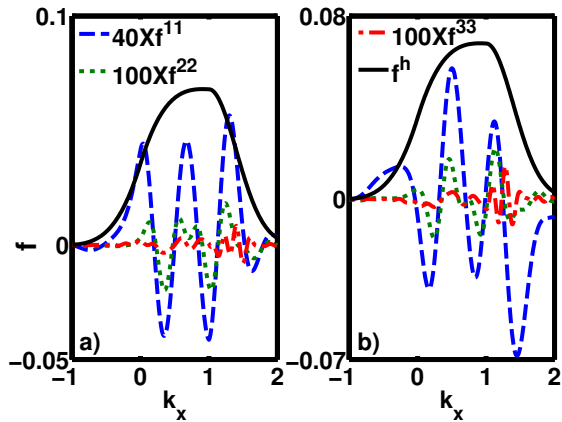}
\caption{(Color online) (a) Cross-sections of real parts of the harmonics $f^{11}$,$f^{22}$,$f^{33}$, and homogeneous steady state distribution function $f^h$. (b)Imaginary parts and $f^h$. Dash ($40Xf^{11}$), dot ($100Xf^{22}$), dash-dot ($100Xf^{33}$), and solid ($f^h$). $F_o = 1kV/cm$, $F_1/F_o = 0.1$, $\omega/\omega_F = 0.5$, $\tau_o/\tau_1 = 0.5$, and $\gamma = 0.01$.}
\vspace{-2.0em}
\label{fig1b}
\end{figure}
Eq. 8 can be solved numerically by discretizing the integral,\cite{Jerri} which leads to a unitary matrix equation of the form $\tilde{K}\tilde{f} = \tilde{f}$. It is found that these eigenvectors exist only for $qv_F = n\omega$, which reduces Eq. 5 to the steady state-homogenous equation for $f^h = f_0^h$.\cite{Sekwao} A similar analysis on the harmonics $f^{m1}, m > 1$ shows that solutions exist only, for $mqv_F  = \omega$, so $f^{m1}$ also reduces to $f_0^h$. Since $f^{mn}$ are related to $f^{(m-1)(n-1)}$ by Eq. 5, it can be deduced from Eq. 5 that solutions for all the harmonics $f^{mn}$ such that $m \neq n > 0$ that exist only when $\omega = mqv_F/n$, yield $f^{mn} = f_0^h$, and should therefore be discarded. As a result, only the harmonics $f^{mn}$ such that $m,n > 0$ and $m = n >0$ are the only remaining terms of the series (4) for $f(\vec k,x,t)$.
%\section{Results}
\par
The current density on the plane is given by,
\begin{equation}J_x(x,t) = -4ev_F\sum_{\vec k'}\Re{f(\vec k',x,t)}\cos(\phi')\end{equation}
where $\phi'$ is the angle between $\vec k'$ and the $k_x'$ axis, and $\Re{f(\vec k',x,t)}$ is the real part of the distribution function.
In our analysis, we write the a-c frequency in units of $\omega/\omega_F$, where $\omega_F = 2\pi eF_ov_F/\hbar \omega_{OP}$, and the wavelength is modulated by a dimensionless parameter $\alpha$ such that
\begin{equation}
 q = \frac{2\pi \alpha}{\lambda_F},
\end{equation}
where $\lambda_F = \hbar\omega_{OP}/eF_o$, and $0 <\alpha\leq 1$. As in a previous work,\cite{Sekwao2} the dimensionless damping parameter $\gamma$ given by
\begin{equation}
\gamma = \frac{1}{\tau_o\sum_{\vec k'}S_{op}(\vec k, \vec k')|_{k = 1.5k_c}}
\end{equation}
is used to modulate the strength of low energy collisions compared to OP scattering.
\par
Figure 2 shows cross sections of the real parts (2a), and imaginary parts (2b) of the first three harmonics $f^{11}$,$f^{22}$,$f^{33}$, with the steady state homogeneous distribution $f^h$ along $k_y = 0$. The applied field in this case is $F_o = 1$kV/cm, such that $F_1/F_o = 0.1$, $\omega/\omega_F = 0.5$, and $\alpha = 1$. The low energy damping parameter is set at $\gamma = 0.01$, and $\tau_o/\tau_1 = 0.5$. From the figure, it can be seen that the harmonics oscillate as a function of $k_x$ with amplitude $f^{mm}$ decreasing as $m$ increases, justifying our choice of the solution (4). We also note that $f$ is always positive as $f^h \gg f^{mn}$.
\par
\begin{figure*}[htp]
\centering
 \includegraphics[width = \linewidth]{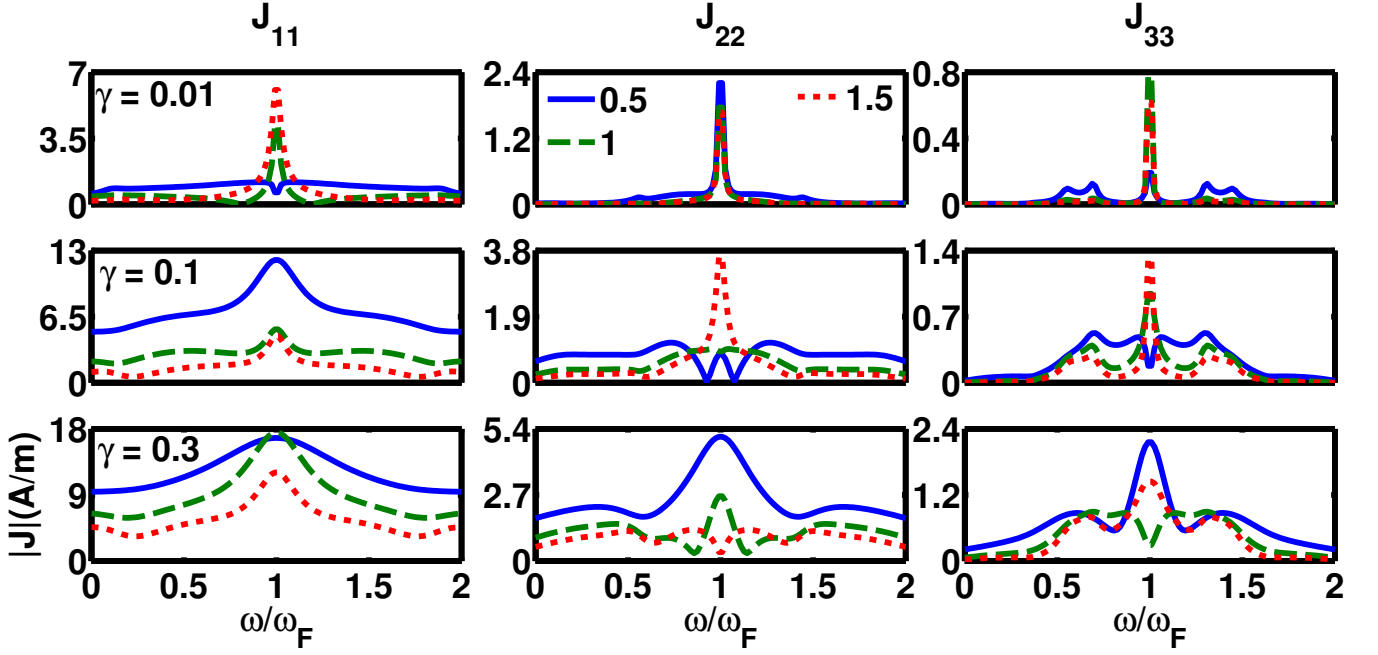}
\caption{(Color online) Current density amplitude versus frequency for different values of $\gamma$ and $F_o$. (a) First harmonic. (b) Second harmonic. (c) Third harmonic. Top row ($\gamma = 0.01$), middle row ($\gamma = 0.1$), and bottom row ($\gamma = 0.3$). Solid ($F_o = 0.5kV/cm$), dash ($F_o = 1kV/cm$), and dot ($F_o = 1.5kV/cm$). $F_1/F_o = 0.1$, $\tau_o/\tau_1 = 0.5$, and $\alpha = 1$.}
\vspace{-2.0em}
\label{fig1b}
\end{figure*}
Figure 3 displays the current density amplitude relative to the first three harmonics as a function of frequency for different values of $F_o$ and $\gamma$. As in Fig. 2, we use $F_1/F_o = 0.1$, $\tau_o/\tau_1 = 0.5$, and $\alpha = 1$. For all three harmonics, there are resonances that occur at $\omega = \omega_F$, for which the amplitudes $f^{mm}(k_x,k_y)$ are real, and the currents are in phase with the applied field. By considering the first row of Fig. 3 ($\gamma = 0.01$), one observes that there is no resonance for $F_o = 0.5$kV/cm (first column). This is because the contribution of the negative values of $f^{11}(k_x,k_y)$ in the total current (Eq. 9) offset that of the positive values at $\omega = \omega_F$. As the field increases from $0.5$kV/cm to $1$kV/cm, more electrons are able to escape low energy scattering in region $I$ and reach the boundary $k = k_c$ and beyond, so the amplitude $f^{11}(k_x,k_y)$ increases in region $II$, inducing a net positive current, so as to achieve resonance. For all three harmonics, the resonance “Quality factor”, $Q = \omega_F/\Delta\omega_{FWHM}$, where $\Delta\omega_{FWHM}$ is the full width of the current density profile at half maximum value, first increases ($Q \approx 24.9$ to $Q \approx 25.2$ for the second, and $Q \approx 23$ to $Q \approx 28$ for the third) as the field increases from $0.5$kV/cm to $1$kV/cm. A further increase in the field to $1.5$kV/cm causes more electrons to reach region $II$ and scatter with OPs, broadening the distribution in the process. As a result, $Q$ decreases (from $Q \approx 20$ to $Q \approx 13$ for the first harmonic, from $Q \approx 25.2$ to $Q \approx 24$ for the second, and from $Q \approx 28$ to $Q \approx 27$ for the third) as the field increase from $1$kV/cm to $1.5$kV/cm.
\par
As low energy scattering ($\gamma$) increases, the system achieves resonance even at low fields (Fig. 3 first column). For the first harmonic, an increase in $\gamma$ makes the amplitude more positive in region $I$, and the overall current becomes positive, achieving resonance. One can also see that the current density amplitude increases, while the overall resonance $Q$- factor decreases as low energy scattering increases. For the first harmonic with $\gamma = 0.1$ (second row, first column), one get $Q \approx 0.7$ ($F_o = 0.5$kV/cm), $Q \approx 0.7$ ($F_o = 1$kV/cm), and $Q \approx 6$ ($F_o = 1.5$kV/cm), only. This is due to the fact that low energy scattering redistribute carriers towards the $\vec k = 0$ region. As seen in a previous work,\cite{Sekwao1} carrier interactions with OPs are essential for resonance to occur. Higher $\gamma$-damping causes fewer carriers to interact with OPs and the resonance $Q$-factor decreases. The current density amplitude increases because the distribution increases around $\vec k = 0$. We note that when $\omega = \omega_F$, the system is a mixture of modes of oscillations with resonant frequencies $m\omega_F$.
\par
\begin{figure}[htp]
\centering
\includegraphics[width = 0.8\linewidth]{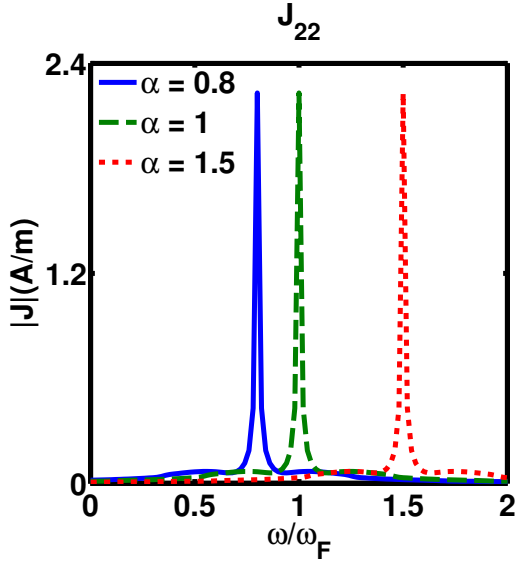}
\caption{(Color online) Second harmonic current density amplitude vs frequency for different values of $\alpha$. Solid ($\alpha = 0.8$), dash ($\alpha = 1.0$) and dot ($\alpha = 1.5$). $F_1/F_o = 0.1$, $\tau_o/\tau_1 = 0.5$, and $\gamma = 0.01$.}
\vspace{-2.0em}
\label{fig1b}
\end{figure}
Figure 4 shows plots of the second harmonic current density versus frequency for different values of $\alpha$ (gate spacing). The damping in this case is again set at $\gamma = 0.01$, $F_1/F_o = 0.1$, and $\tau_o/\tau_1 = 0.5$. From the figure, the observed resonances are achieved at the frequencies $\omega_F$, $0.8\omega_F$, and $1.5\omega_F$ corresponding to $\alpha = 1$, $\alpha = 0.8$, and $\alpha = 1.5$ respectively, which shows that for a particular $F_o$, the resonance frequency is tunable with the wavelength of the applied field. This important result indicates the potential use of graphene in terahertz sources and detectors.\cite{Sekwao1,Sekwao2}
\par
In conclusion, we have provided an analysis of carrier dynamics in graphene subjected to periodic, time, and space dependent electric fields and scattering times. Our model shows that, high $Q$ resonances can be achieved when $\omega = \omega_F$. As expected, the $Q$-factors decrease with damping $\gamma$. We also observe that at resonance, the system consists of carrier excitations with frequencies $m\omega_F$, for $m \geq 1$. As a result the system is essentially a mixer since an input frequency $\omega_F$, creates the harmonics $m\omega_F$, and appropriate filters should be used to pick out the required frequencies. Also, the resonant frequency is tunable with the wavelength of the applied field. Note that the wavelength $\lambda$ of the a-c field is twice the distance between successive gates in Fig. 1. Consequently, graphene can potentially be used to make high power, tunable terahertz devices that operate at room temperature.
\begin{acknowledgments}
This work was supported by ARO Grant No. W911NF-11-1-0434. The authors gratefully acknowledge the use of the Taub cluster maintained and operated by the Computational Science and Engineering Program at the University of Illinois.
\end{acknowledgments}
\bibliography{mybib}

\begin{thebibliography}{18}%
\makeatletter
\providecommand \@ifxundefined [1]{%
 \@ifx{#1\undefined}
}%
\providecommand \@ifnum [1]{%
 \ifnum #1\expandafter \@firstoftwo
 \else \expandafter \@secondoftwo
 \fi
}%
\providecommand \@ifx [1]{%
 \ifx #1\expandafter \@firstoftwo
 \else \expandafter \@secondoftwo
 \fi
}%
\providecommand \natexlab [1]{#1}%
\providecommand \enquote  [1]{``#1''}%
\providecommand \bibnamefont  [1]{#1}%
\providecommand \bibfnamefont [1]{#1}%
\providecommand \citenamefont [1]{#1}%
\providecommand \href@noop [0]{\@secondoftwo}%
\providecommand \href [0]{\begingroup \@sanitize@url \@href}%
\providecommand \@href[1]{\@@startlink{#1}\@@href}%
\providecommand \@@href[1]{\endgroup#1\@@endlink}%
\providecommand \@sanitize@url [0]{\catcode `\\12\catcode `\$12\catcode
  `\&12\catcode `\#12\catcode `\^12\catcode `\_12\catcode `\%12\relax}%
\providecommand \@@startlink[1]{}%
\providecommand \@@endlink[0]{}%
\providecommand \url  [0]{\begingroup\@sanitize@url \@url }%
\providecommand \@url [1]{\endgroup\@href {#1}{\urlprefix }}%
\providecommand \urlprefix  [0]{URL }%
\providecommand \Eprint [0]{\href }%
\providecommand \doibase [0]{http://dx.doi.org/}%
\providecommand \selectlanguage [0]{\@gobble}%
\providecommand \bibinfo  [0]{\@secondoftwo}%
\providecommand \bibfield  [0]{\@secondoftwo}%
\providecommand \translation [1]{[#1]}%
\providecommand \BibitemOpen [0]{}%
\providecommand \bibitemStop [0]{}%
\providecommand \bibitemNoStop [0]{.\EOS\space}%
\providecommand \EOS [0]{\spacefactor3000\relax}%
\providecommand \BibitemShut  [1]{\csname bibitem#1\endcsname}%
\let\auto@bib@innerbib\@empty
%</preamble>
\bibitem [{\citenamefont {Novoselov}\ \emph {et~al.}(2005)\citenamefont
  {Novoselov}, \citenamefont {Geim}, \citenamefont {Morozov}, \citenamefont
  {Jiang}, \citenamefont {Katsnelson}, \citenamefont {Grigorieva},
  \citenamefont {Dubonos},\ and\ \citenamefont {Fristov}}]{Novoselov}%
  \BibitemOpen
  \bibfield  {author} {\bibinfo {author} {\bibfnamefont {K.~S.}\ \bibnamefont
  {Novoselov}}, \bibinfo {author} {\bibfnamefont {A.~K.}\ \bibnamefont {Geim}},
  \bibinfo {author} {\bibfnamefont {S.~V.}\ \bibnamefont {Morozov}}, \bibinfo
  {author} {\bibfnamefont {D.}~\bibnamefont {Jiang}}, \bibinfo {author}
  {\bibfnamefont {M.~I.}\ \bibnamefont {Katsnelson}}, \bibinfo {author}
  {\bibfnamefont {I.~V.}\ \bibnamefont {Grigorieva}}, \bibinfo {author}
  {\bibfnamefont {S.~V.}\ \bibnamefont {Dubonos}}, \ and\ \bibinfo {author}
  {\bibfnamefont {A.~A.}\ \bibnamefont {Fristov}},\ }\href@noop {} {\bibfield
  {journal} {\bibinfo  {journal} {Nature (London)}\ }\textbf {\bibinfo {volume}
  {438}},\ \bibinfo {pages} {197} (\bibinfo {year} {2005})}\BibitemShut
  {NoStop}%
\bibitem [{\citenamefont {Novoselov}\ and\ \citenamefont
  {Geim}(2007)}]{Novoselov2}%
  \BibitemOpen
  \bibfield  {author} {\bibinfo {author} {\bibfnamefont {K.~S.}\ \bibnamefont
  {Novoselov}}\ and\ \bibinfo {author} {\bibfnamefont {A.~K.}\ \bibnamefont
  {Geim}},\ }\href@noop {} {\bibfield  {journal} {\bibinfo  {journal} {Nature
  Materials}\ }\textbf {\bibinfo {volume} {6}},\ \bibinfo {pages} {183}
  (\bibinfo {year} {2007})}\BibitemShut {NoStop}%
\bibitem [{\citenamefont {Katsnelson}(2007)}]{Katsnelson}%
  \BibitemOpen
  \bibfield  {author} {\bibinfo {author} {\bibfnamefont {M.~I.}\ \bibnamefont
  {Katsnelson}},\ }\href@noop {} {\bibfield  {journal} {\bibinfo  {journal}
  {Materials Today}\ }\textbf {\bibinfo {volume} {10}},\ \bibinfo {pages} {20}
  (\bibinfo {year} {2007})}\BibitemShut {NoStop}%
\bibitem [{\citenamefont {Saito}\ \emph {et~al.}(1998)\citenamefont {Saito},
  \citenamefont {Dresselhaus},\ and\ \citenamefont {Dresselhaus}}]{Saito}%
  \BibitemOpen
  \bibfield  {author} {\bibinfo {author} {\bibfnamefont {R.}~\bibnamefont
  {Saito}}, \bibinfo {author} {\bibfnamefont {G.}~\bibnamefont {Dresselhaus}},
  \ and\ \bibinfo {author} {\bibfnamefont {M.~S.}\ \bibnamefont
  {Dresselhaus}},\ }\href@noop {} {\emph {\bibinfo {title} {Physical Properties
  of Carbon Nanotubes}}}\ (\bibinfo  {publisher} {Imperial College Press},\
  \bibinfo {year} {1998})\BibitemShut {NoStop}%
\bibitem [{\citenamefont {Mikhailov}(2007)}]{Mikhailov}%
  \BibitemOpen
  \bibfield  {author} {\bibinfo {author} {\bibfnamefont {S.~A.}\ \bibnamefont
  {Mikhailov}},\ }\href@noop {} {\bibfield  {journal} {\bibinfo  {journal}
  {Europhysics Letters}\ }\textbf {\bibinfo {volume} {79}},\ \bibinfo {pages}
  {27002} (\bibinfo {year} {2007})}\BibitemShut {NoStop}%
\bibitem [{\citenamefont {Vafek}(2006)}]{Vafek}%
  \BibitemOpen
  \bibfield  {author} {\bibinfo {author} {\bibfnamefont {O.}~\bibnamefont
  {Vafek}},\ }\href@noop {} {\bibfield  {journal} {\bibinfo  {journal}
  {Physical Review Letters}\ }\textbf {\bibinfo {volume} {97}},\ \bibinfo
  {pages} {266406} (\bibinfo {year} {2006})}\BibitemShut {NoStop}%
\bibitem [{\citenamefont {Ryzhii}\ \emph {et~al.}(2007)\citenamefont {Ryzhii},
  \citenamefont {Satou},\ and\ \citenamefont {Otsuji}}]{Ryzhii}%
  \BibitemOpen
  \bibfield  {author} {\bibinfo {author} {\bibfnamefont {V.}~\bibnamefont
  {Ryzhii}}, \bibinfo {author} {\bibfnamefont {A.}~\bibnamefont {Satou}}, \
  and\ \bibinfo {author} {\bibfnamefont {T.}~\bibnamefont {Otsuji}},\
  }\href@noop {} {\bibfield  {journal} {\bibinfo  {journal} {Journal of Applied
  Physics}\ }\textbf {\bibinfo {volume} {101}},\ \bibinfo {pages} {024509}
  (\bibinfo {year} {2007})}\BibitemShut {NoStop}%
\bibitem [{\citenamefont {Vasko}\ and\ \citenamefont {Ryzhii}(2008)}]{Vasko}%
  \BibitemOpen
  \bibfield  {author} {\bibinfo {author} {\bibfnamefont {F.~T.}\ \bibnamefont
  {Vasko}}\ and\ \bibinfo {author} {\bibfnamefont {V.}~\bibnamefont {Ryzhii}},\
  }\href@noop {} {\bibfield  {journal} {\bibinfo  {journal} {Physical Review
  B}\ }\textbf {\bibinfo {volume} {77}},\ \bibinfo {pages} {195433} (\bibinfo
  {year} {2008})}\BibitemShut {NoStop}%
\bibitem [{\citenamefont {Piscanec}\ \emph {et~al.}(2007)\citenamefont
  {Piscanec}, \citenamefont {Lazzeri}, \citenamefont {Mauri},\ and\
  \citenamefont {Ferrari}}]{Piscanec}%
  \BibitemOpen
  \bibfield  {author} {\bibinfo {author} {\bibfnamefont {S.}~\bibnamefont
  {Piscanec}}, \bibinfo {author} {\bibfnamefont {M.}~\bibnamefont {Lazzeri}},
  \bibinfo {author} {\bibfnamefont {F.}~\bibnamefont {Mauri}}, \ and\ \bibinfo
  {author} {\bibfnamefont {A.}~\bibnamefont {Ferrari}},\ }\href@noop {}
  {\bibfield  {journal} {\bibinfo  {journal} {Eur. Phys. J. Special Topics}\
  }\textbf {\bibinfo {volume} {148}},\ \bibinfo {pages} {159} (\bibinfo {year}
  {2007})}\BibitemShut {NoStop}%
\bibitem [{\citenamefont {Ferrari}\ \emph {et~al.}(2006)\citenamefont
  {Ferrari}, \citenamefont {Meyer}, \citenamefont {Scardaci}, \citenamefont
  {Casiraghi}, \citenamefont {Lazzeri}, \citenamefont {Mauri}, \citenamefont
  {Piscanec}, \citenamefont {Jiang}, \citenamefont {Novoselov}, \citenamefont
  {Roth},\ and\ \citenamefont {Geim}}]{Ferrari}%
  \BibitemOpen
  \bibfield  {author} {\bibinfo {author} {\bibfnamefont {A.~C.}\ \bibnamefont
  {Ferrari}}, \bibinfo {author} {\bibfnamefont {J.~C.}\ \bibnamefont {Meyer}},
  \bibinfo {author} {\bibfnamefont {V.}~\bibnamefont {Scardaci}}, \bibinfo
  {author} {\bibfnamefont {C.}~\bibnamefont {Casiraghi}}, \bibinfo {author}
  {\bibfnamefont {M.}~\bibnamefont {Lazzeri}}, \bibinfo {author} {\bibfnamefont
  {F.}~\bibnamefont {Mauri}}, \bibinfo {author} {\bibfnamefont
  {S.}~\bibnamefont {Piscanec}}, \bibinfo {author} {\bibfnamefont
  {D.}~\bibnamefont {Jiang}}, \bibinfo {author} {\bibfnamefont {K.~S.}\
  \bibnamefont {Novoselov}}, \bibinfo {author} {\bibfnamefont {S.}~\bibnamefont
  {Roth}}, \ and\ \bibinfo {author} {\bibfnamefont {A.~K.}\ \bibnamefont
  {Geim}},\ }\href@noop {} {\bibfield  {journal} {\bibinfo  {journal} {Physical
  Review Letters}\ }\textbf {\bibinfo {volume} {97}},\ \bibinfo {pages}
  {187401} (\bibinfo {year} {2006})}\BibitemShut {NoStop}%
\bibitem [{\citenamefont {Tiwari}(1992)}]{Tiwari}%
  \BibitemOpen
  \bibfield  {author} {\bibinfo {author} {\bibfnamefont {S.}~\bibnamefont
  {Tiwari}},\ }\href@noop {} {\emph {\bibinfo {title} {Compound Semiconductor
  Device Physics}}}\ (\bibinfo  {publisher} {Academic Press},\ \bibinfo
  {address} {San Diego, CA},\ \bibinfo {year} {1992})\BibitemShut {NoStop}%
\bibitem [{\citenamefont {Shockley}(1951)}]{Shockley}%
  \BibitemOpen
  \bibfield  {author} {\bibinfo {author} {\bibfnamefont {W.}~\bibnamefont
  {Shockley}},\ }\href@noop {} {\bibfield  {journal} {\bibinfo  {journal} {Bell
  System Technical Journal}\ }\textbf {\bibinfo {volume} {30}},\ \bibinfo
  {pages} {990} (\bibinfo {year} {1951})}\BibitemShut {NoStop}%
\bibitem [{\citenamefont {Sekwao}\ and\ \citenamefont
  {Leburton}(2011)}]{Sekwao}%
  \BibitemOpen
  \bibfield  {author} {\bibinfo {author} {\bibfnamefont {S.}~\bibnamefont
  {Sekwao}}\ and\ \bibinfo {author} {\bibfnamefont {J.-P.}\ \bibnamefont
  {Leburton}},\ }\href@noop {} {\bibfield  {journal} {\bibinfo  {journal}
  {Physical Review B}\ }\textbf {\bibinfo {volume} {83}},\ \bibinfo {pages}
  {075418} (\bibinfo {year} {2011})}\BibitemShut {NoStop}%
\bibitem [{\citenamefont {Neto}\ \emph {et~al.}(2009)\citenamefont {Neto},
  \citenamefont {Guinea}, \citenamefont {Peres}, \citenamefont {Novoselov},\
  and\ \citenamefont {Geim}}]{Castro}%
  \BibitemOpen
  \bibfield  {author} {\bibinfo {author} {\bibfnamefont {A.~H.~C.}\
  \bibnamefont {Neto}}, \bibinfo {author} {\bibfnamefont {F.}~\bibnamefont
  {Guinea}}, \bibinfo {author} {\bibfnamefont {N.~M.~R.}\ \bibnamefont
  {Peres}}, \bibinfo {author} {\bibfnamefont {K.~S.}\ \bibnamefont
  {Novoselov}}, \ and\ \bibinfo {author} {\bibfnamefont {A.~K.}\ \bibnamefont
  {Geim}},\ }\href@noop {} {\bibfield  {journal} {\bibinfo  {journal} {Review
  of Modern Physics}\ }\textbf {\bibinfo {volume} {81}},\ \bibinfo {pages}
  {109} (\bibinfo {year} {2009})}\BibitemShut {NoStop}%
\bibitem [{\citenamefont {Lazzeri}\ \emph {et~al.}(2005)\citenamefont
  {Lazzeri}, \citenamefont {Piscanec}, \citenamefont {Mauri}, \citenamefont
  {Ferrari},\ and\ \citenamefont {Robertson}}]{Lazzeri}%
  \BibitemOpen
  \bibfield  {author} {\bibinfo {author} {\bibfnamefont {M.}~\bibnamefont
  {Lazzeri}}, \bibinfo {author} {\bibfnamefont {S.}~\bibnamefont {Piscanec}},
  \bibinfo {author} {\bibfnamefont {F.}~\bibnamefont {Mauri}}, \bibinfo
  {author} {\bibfnamefont {A.~C.}\ \bibnamefont {Ferrari}}, \ and\ \bibinfo
  {author} {\bibfnamefont {J.}~\bibnamefont {Robertson}},\ }\href@noop {}
  {\bibfield  {journal} {\bibinfo  {journal} {Physical Review Letters}\
  }\textbf {\bibinfo {volume} {95}},\ \bibinfo {pages} {236802} (\bibinfo
  {year} {2005})}\BibitemShut {NoStop}%
\bibitem [{\citenamefont {Jerri}(1999)}]{Jerri}%
  \BibitemOpen
  \bibfield  {author} {\bibinfo {author} {\bibfnamefont {A.~J.}\ \bibnamefont
  {Jerri}},\ }\href@noop {} {\emph {\bibinfo {title} {Introduction to Integral
  Equations with Applications}}}\ (\bibinfo  {publisher} {John Wiley and
  Sons},\ \bibinfo {year} {1999})\BibitemShut {NoStop}%
\bibitem [{\citenamefont {Sekwao}\ and\ \citenamefont
  {Leburton}(2013{\natexlab{a}})}]{Sekwao2}%
  \BibitemOpen
  \bibfield  {author} {\bibinfo {author} {\bibfnamefont {S.}~\bibnamefont
  {Sekwao}}\ and\ \bibinfo {author} {\bibfnamefont {J.-P.}\ \bibnamefont
  {Leburton}},\ }\href@noop {} {\bibfield  {journal} {\bibinfo  {journal}
  {Applied Physics Letters}\ }\textbf {\bibinfo {volume} {103}},\ \bibinfo
  {pages} {143108} (\bibinfo {year} {2013}{\natexlab{a}})}\BibitemShut
  {NoStop}%
\bibitem [{\citenamefont {Sekwao}\ and\ \citenamefont
  {Leburton}(2013{\natexlab{b}})}]{Sekwao1}%
  \BibitemOpen
  \bibfield  {author} {\bibinfo {author} {\bibfnamefont {S.}~\bibnamefont
  {Sekwao}}\ and\ \bibinfo {author} {\bibfnamefont {J.-P.}\ \bibnamefont
  {Leburton}},\ }\href@noop {} {\bibfield  {journal} {\bibinfo  {journal}
  {Physical Review B}\ }\textbf {\bibinfo {volume} {87}},\ \bibinfo {pages}
  {155424} (\bibinfo {year} {2013}{\natexlab{b}})}\BibitemShut {NoStop}%
\end{thebibliography}%
\end{document}